\documentclass[iop]{emulateapj}
\usepackage{graphicx}
\usepackage{natbib}
\usepackage{txfonts}
\begin{document} 

\title{Sunquake generation by coronal magnetic restructuring}

\author{A. J. B. Russell and M. K. Mooney\altaffilmark{1}}
\affil{School of Science \& Engineering, University of Dundee, Dundee, DD1 4HN, Scotland, U.K.}

\author{ J. E. Leake}
\affil{Naval Research Laboratory, Washington, DC 20375, U.S.A.}
\author{H. S. Hudson\altaffilmark{2}}
\affil{Space Sciences Lab, University of California Berkeley, Berkeley, CA 94720, U.S.A.}

\altaffiltext{1}{Also at the School of Physics and Astronomy, University of St. Andrews, St Andrews, KY16 9SS, Scotland, U.K.}
\altaffiltext{2}{SUPA School of Physics and Astronomy, University of Glasgow, Glasgow, G12 8QQ, Scotland, U.K.}

\shorttitle{Sunquake generation}
\shortauthors{Russell et al.}

\begin{abstract}
   Sunquakes are the surface signatures of acoustic waves in the Sun's interior that are produced by some but not all flares and coronal mass ejections (CMEs).
   This paper explores a mechanism for sunquake generation by the magnetic field changes that occur during flares and CMEs,
   using MHD simulations with a semiempirical FAL-C atmosphere to demonstrate the generation of acoustic waves in the interior in response to changing magnetic tilt in the corona.  
   We find that Alfv\'en-sound resonance combined with the ponderomotive force produces acoustic waves in the interior with sufficient energy to match sunquake observations when the magnetic field angle changes by the order of 10 degrees in a region where the coronal field strength is a few hundred gauss or more.  
   The most energetic sunquakes are produced when the coronal field is strong,
   while the variation of magnetic field strength with height and the time scale of the tilt change are of secondary importance.
\end{abstract}


\maketitle

\section{Introduction}\label{sec:intro}
Sunquakes are seismic waves that are observed for some but not all coronal mass ejections (CMEs) and M and X class flares.  They were first detected on the Sun by \citet{1998KosivichevZharkova} and have since been observed many times, e.g. the events listed by \citet{2011Donea}.  The associated acoustic wave typically has an energy between $10^{27}$ and  $10^{29}\mathrm{~erg}$ and comes from a source with an area on the order of $10\mathrm{~Mm}^2$, implying energy fluences (time-integrated energy fluxes) of $10^{10}$--$10^{12}\mathrm{~erg~cm}^{-2}$. 
There are many open questions about sunquakes, most notably the nature of the excitation mechanism or mechanisms.
The possibilities currently under consideration can be divided into two categories depending on the force that provides the impulse.  

In the first type of mechanism, a pressure wave is generated by impulsive heating \citep{1972Wolff},
which is attributed to thick-target heating of the chromosphere by energetic electrons \citep{1998KosivichevZharkova} or heating of the photosphere due to backwarming \citep{2008LindseyDonea} or deeply penetrating protons \citep{2007ZharkovaZharkov}. Wave heating of the photosphere and chromosphere \citep{2013RussellFletcher,2016ReepRussell} could also come into this class, as suggested by the observations of \citet{2015Matthews}.  These explanations seem particularly suited to events where one or more seismic sources are collocated with hard X-ray or white light sources, respectively indicating energetic electrons in the chromosphere or heating of the photosphere. The main theoretical difficulty is that these mechanisms typically invoke the passage of a shock wave into the interior, and radiative losses are expected to strongly damp such shocks, depleting the energy available for the seismic wave.  It has, however, been suggested by \citet{2014Lindsey} that  horizontal magnetic field could reduce the radiative losses.

The other driver is Lorentz forces, first suggested by \citet{2008Hudson} and refined by \citet{2012Fisher}.  Solar flares and coronal mass ejections both involve extensive coronal magnetic restructuring and there is convincing evidence that this changes the photospheric magnetic field.  Care is needed when interpreting spectropolarimetric data during a flare since the plasma may not be in local thermal equilibrium, however, photospheric magnetic fields do seem to make rapid irreversible changes during the flare impulsive phase \citep{1994Wang,2005SudolHarvey,2010WangLiu,2010PetrieSudol}
including abrupt changes to the direction of the magnetic field by several tens of degrees \citep{2013Petrie}, 
often in association with increased UV emission from the overlying chromosphere \citep{2012Johnstone}.
Lorentz forces seem particularly relevant to sunquakes where a magnetic change was seen at a seismic source \citep[e.g.][]{2011Kumar} or where seismic source locations correspond to footpoints of an erupting flux rope  \citep{2011Zharkov}.

This paper presents the first MHD simulations of sunquake generation by Lorentz forces.
This new approach to sunquakes yields significant advances in understanding, 
allowing us to give the first complete account of how magnetic changes launch the acoustic wave,
and showing for the first time that a realistic change to the magnetic field in the corona produces an acoustic wave with sufficient energy flux to match sunquake observations.

\begin{figure*}
\centering
\centering
\includegraphics[width=\hsize]{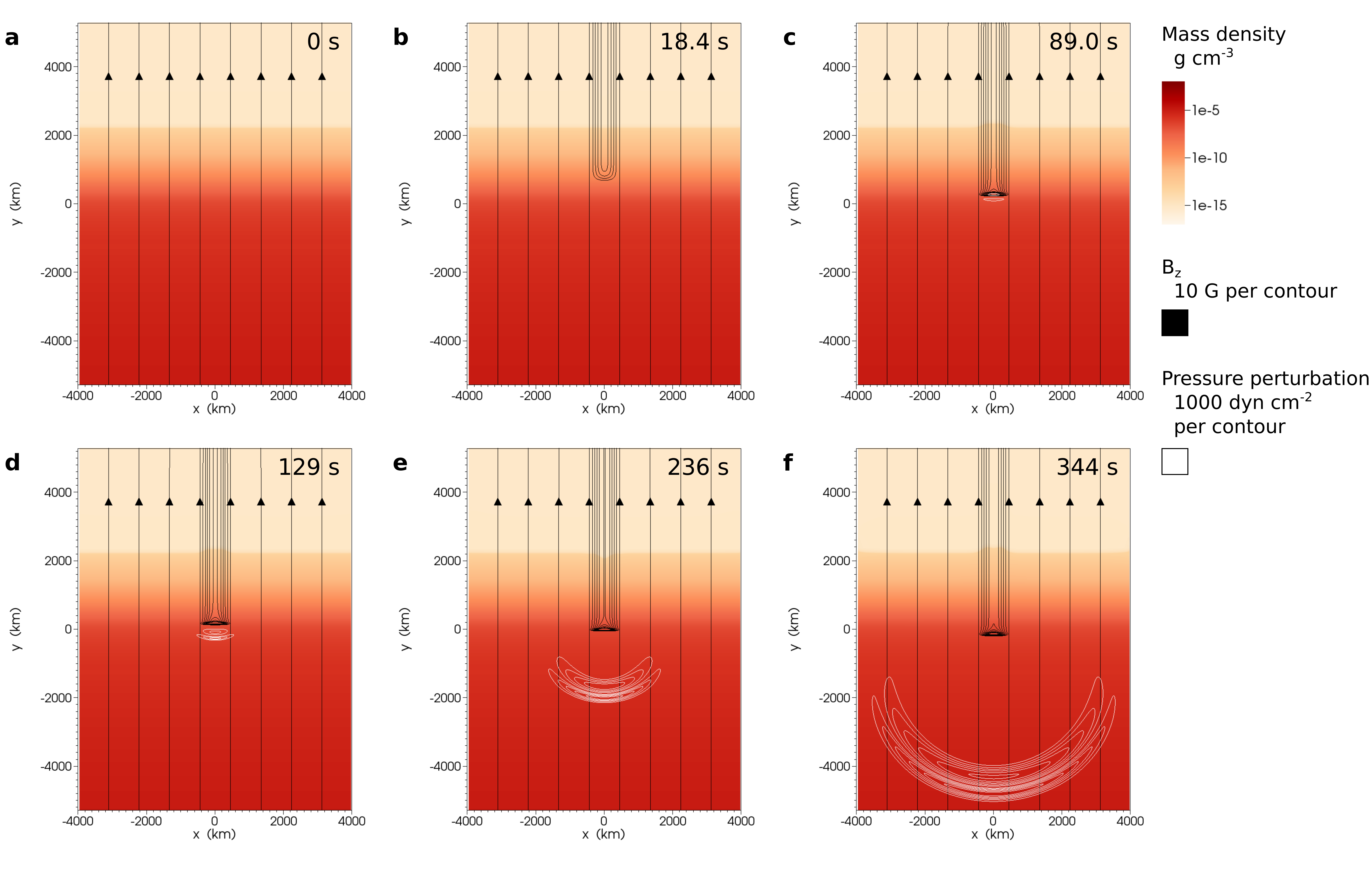}
\caption{A 2D MHD simulation of acoustic wave generation by changing coronal magnetic field. 
The logarithmic colour scale shows the mass density, field lines show the magnetic field in the $x$-$y$ plane (initially 250 G), black contours (extending from the top boundary) show $B_z$, and white contours (the double bow shapes) show the perturbed plasma pressure, $\delta p(\mathbf{x},t) = p(\mathbf{x},t)-p_0(\mathbf{x})$. The transition region is the sharp change in density between 2200 and 2400~km and the sound and Alfv\'en speeds match at 452~km.  The leading pressure perturbation contains increased pressure and the following one has decreased pressure.
This figure is also available as an animation in the electronic edition of the journal.}\label{fig:2d}
\end{figure*}

\section{Simulation setup}\label{sec:model}
Our study uses an initial model solar atmosphere that extends from the interior to the corona. The semiempirical FAL-C model by \citet{2006Fontenla} provides temperatures for the chromosphere, and we extend the model upwards by joining a 1~MK corona to the top of the FAL-C model with a 150~km thick upper transition region between them, and downwards into the interior using a linear function of height that represents an adiabatic polytrope. Densities are then obtained under an assumption of hydrostatic equilibrium with the photospheric density prescribed by the FAL-C model. The final component is a magnetic field, which is assumed vertical in this work since that allows the simplest demonstration of the proposed mechanism. The initial equilibrium is shown in Fig.~\ref{fig:2d}(\textbf{a}).
 
We evolve the model by solving the standard nonlinear magnetohydrodynamic (MHD) equations in 2.5 dimensions using the Lare2d code \citep[the equations and numerical details can be found in][]{2001Arber}.  2.5D means that invariance is assumed in one of the horizontal directions, which we take as the $z$ coordinate with $y$ the vertical coordinate and $x$ completing the Cartesian triad.  The time scales of interest are long compared to the neutral-ion coupling time in the chromosphere (at most a few tens of milliseconds) so a single fluid description is appropriate.  Resistivity, viscosity, thermal conduction and radiation are neglected for simplicity since they are not essential for sunquake generation to occur, although we caution that these effects may reduce the acoustic energy when they are ultimately included.  As discussed in Sec.~\ref{sec:disc}, future work should examine the impact of radiation on the acoustic waves, while heating due to Cowling resistivity merits investigation in the context of white light emissions at sunquake sources.  

The simulation is driven by imposing a velocity in the invariant $z$ direction in the hot corona, which smoothly and continuously displaces the plasma over a chosen time scale, $\tau_d$, thereby tilting the magnetic field.  The driving is applied at the top boundary of the simulation, $y=5286\textrm{~km}$.

\section{Sunquake generation}\label{sec:2d}
Figure \ref{fig:2d} shows a simulation where the coronal magnetic field tilts during a 30~s interval, reaching a maximum inclination of 11 degrees (measured at the top of the domain).  The initial field strength is 250~G.

\begin{figure}
\centering
\includegraphics[height=10cm]{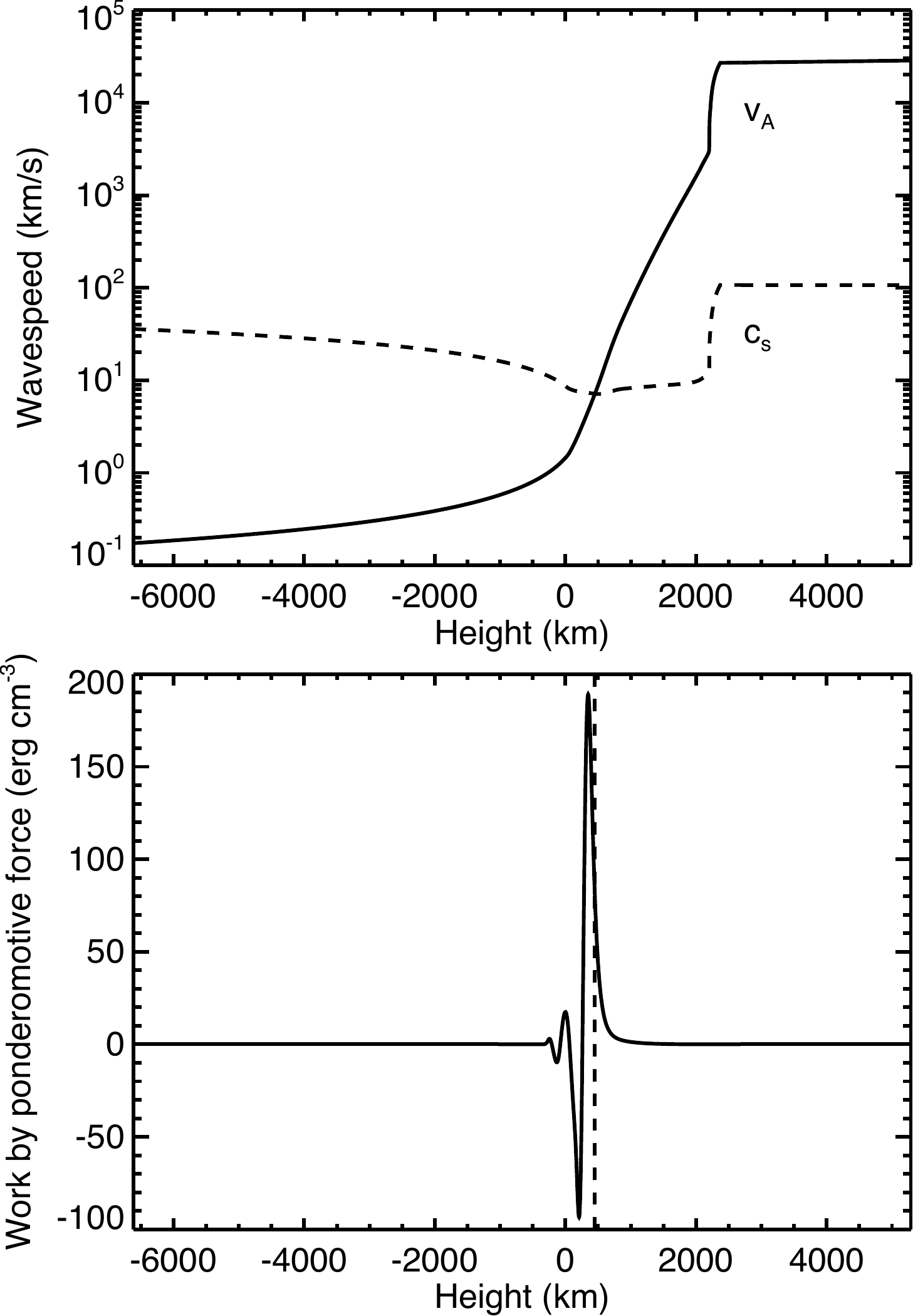}
\caption{Top: Alfv\'en speed $v_A$ and sound speed $c_s$ as functions of height for the initial FAL-C equilibrium with $B_0$ of 250~G. The speeds are equal at 452 km.  Bottom: Work done by the ponderomotive force in a 1D simulation of sunquake generation, showing a resonant peak where the wave speeds match.}\label{fig:resonance}
\end{figure}

The black contours in Fig.~\ref{fig:2d}(\textbf{b}) show $B_z$ at ${t=18.4\mathrm{~s}}$ (measured from the start of the driving). The field change is localized in $x$ with a finite extent imposed by the driver. In the coronal part of the model, the magnetic field is essentially invariant with height -- a consequence of the large coronal Alfv\'en speed (see Figure \ref{fig:resonance} top) -- however, a wave front for the magnetic field change can be identified in the chromosphere as horizontally-aligned contours of $B_z$ and this propagates as an Alfv\'en wave. As the magnetic change propagates deeper into the atmosphere (Fig.~\ref{fig:2d}(\textbf{c})), the maximum value of $B_z$ in the wave front increases, which we explain as the flux of $B_z$ piling up due to the Alfv\'en speed being significantly slower ahead of the wave front than behind it.
The main time of interest is when the Alfv\'en wave front crosses the equipartition height at which the Alfv\'en speed, $v_A=B/\sqrt{4\pi\rho}$, and the sound speed, $c_s=\sqrt{\gamma k T/m}$, are equal: $y=452~\mathrm{km}$. During this phase an acoustic wave is generated, the start of which is visible in Fig.~\ref{fig:2d}(c) as an increase in the local pressure slightly ahead of the Alfv\'enic front followed by a pressure decrease (white contours).  
The coupling ends when the resonance is lost, after which the sound wave propagates into the interior and refracts due to the stratified sound speed (panels \textbf{d}-\textbf{f}).  Ultimately, the sound wave will intersect the photosphere, creating the sunquake signature. The Alfv\'en wave, meanwhile, has slowed to a virtual standstill on the simulation time scale, so the magnetic field change and associated currents are almost static at later times and do not penetrate far below the photosphere.

We now examine the coupling process.  Writing equilibrium quantities with a subscript 0 and perturbed quantities with a delta prefix, the Lorentz force can be expanded as 
\begin{eqnarray}
\mathbf{F}_L = \mathbf{J}\times\mathbf{B} =  \mathbf{\delta J}\times\mathbf{B}_0+\mathbf{\delta J}\times\mathbf{\delta B}
\end{eqnarray}
(since our initial equilibrium is current-free we do not include $\mathbf{J}_0$ terms).
For an Alfv\'en wave, the leading order $ \mathbf{\delta J}\times\mathbf{B}_0$ term provides the restoring force: it is in the invariant direction, does not compress the plasma and therefore does not couple to the sound wave.  The $\mathbf{\delta J}\times\mathbf{\delta B}$ term is the ponderomotive force: it does have a component parallel to the background magnetic field, which allows coupling to the acoustic mode.  Ponderomotive effects have been widely studied in the context of other solar phenomena, for example, as a potential source of shocks that heat the chromosphere and form spicules \citep{1982Hollweg} and as an explanation for the FIP effect \citep{2015Laming}.
The effects of the ponderomotive force are often small even for nonlinear Alfv\'en waves because the energy transferred from the Alfv\'en wave depends on the scalar product of the ponderomotive force with the plasma velocity along the field.  Coupling is therefore only significant energetically when the growing sound wave (produced by the coupling) is resonant with the Alfv\'en wave ($v_A\approx c_s$).

The top panel of Fig.~\ref{fig:resonance} shows $v_A$ and $c_s$ in the initial equilibrium.  They are equal at 452 km, near which the Alfv\'en speed decreases rapidly with depth while the sound speed varies only slowly.
The bottom panel shows the work done by the ponderomotive force (the time integral of $\mathbf{v}\cdot\mathbf{\delta J}\times\mathbf{\delta B}$ over the entire simulation) at every height for a 1D version of the simulation shown in Fig.~\ref{fig:2d} (introducing the additional assumption of invariance in $x$).
A strong peak is seen at the Alfv\'en-sound resonance,
with the maximum coupling occurring slightly to the lower side of the resonance where $v_y$ is more developed.
There is an anti-resonance below this where the acoustic wave loses energy because it has become out of phase with the Alfv\'en wave, hence work is done against the ponderomotive force, not by it.  Rapidly diminishing resonances and anti-resonances occur lower down.  The height-integrated work transfers $2.4\times10^9\mbox{~erg~cm}^{-2}$ to the acoustic wave, matching the acoustic energy evaluated in Sec.~\ref{sec:energy}, with the main contribution coming from the highest resonant peak.  We conclude that the ponderomotive force and Alfv\'en-sound resonance launch the acoustic wave into the interior.

\section{Energies}\label{sec:energy}
\begin{figure}\centering
\includegraphics[height=10cm]{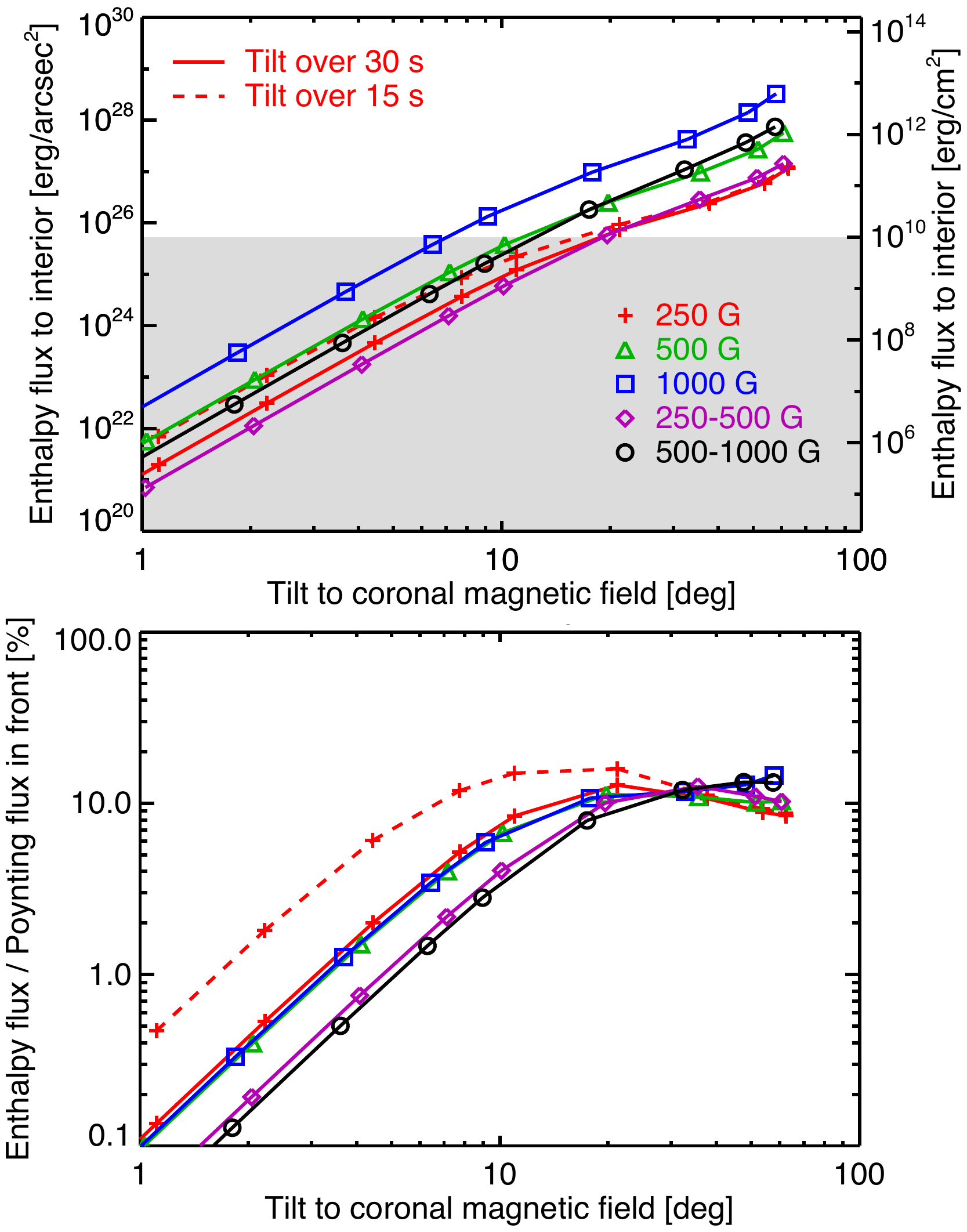}
\caption{Top: Time-integrated enthalpy flux in the acoustic wave (approximately its total energy per source area) for 1D simulations with various magnetic tilts, initial magnetic fields and driving time scales. Bottom: Percentage of the energy in the magnetic front converted to acoustic energy. Where a single magnetic field strength is indicated, the magnetic field was uniform; where a range is given, the field strength was a function of height and the values state the field strength in the corona and at the photosphere.}\label{fig:energies}
\end{figure}

We now investigate whether this mechanism produces acoustic waves with sufficient energy to explain sunquake observations.  

The energy of an acoustic wave with low Mach number is closely approximated by the time-integrated enthalpy flux below the coupling region.  The enthalpy flux \citep[e.g.][]{2009Birn} is
\begin{eqnarray}
 \mathbf{H} = \left(\frac{\gamma}{\gamma-1}\right)p\mathbf{v} 
  = \left(\frac{\gamma}{\gamma-1}\right)p_0\mathbf{\delta v} 
  + \left(\frac{\gamma}{\gamma-1}\right)\delta p\mathbf{\delta v}
\end{eqnarray}
(the $\mathbf{v}_0$ terms vanish for our static initial equilibrium).
Since the passing of an acoustic wave does not produce a net displacement of the plasma, the linear contribution integrates to zero and we integrate only the second order term.

Figure \ref{fig:energies} (top) shows the time-integrated enthalpy flux at ${-2000~\textrm{km}}$ for a collection of 1D simulations.  Points above the shaded region exceed the lower observable limit of $10^{10}\mathrm{~erg~cm}^{-2}$ noted in Sec.~\ref{sec:intro}.  
The solid curve with crosses (red) shows data for a field strength of {250~}G and tilts applied over {30~s}, as considered in Sec.~\ref{sec:2d}. For these parameters, tilts of 20 degrees or more produce acoustic waves that are in principle strong enough to produce observable sunquakes. 
Figure \ref{fig:energies} (bottom) shows the coupling efficiency, defined as the time-integrated enthalpy flux at ${-2000~\textrm{km}}$ divided by the time-integrated Poynting flux at {2000~km} associated with the main Alfv\'enic front.  
This figure shows that there are two regimes.  
Writing $\theta$ for the angle change, for tilts up to around 10 degrees the downgoing Alfv\'enic front evolves as a linear Alfv\'en wave, the Poynting flux in the front varies as $\theta^2$ and the coupling efficiency varies as $\theta^2$, hence the energy of the acoustic wave varies as $\theta^4$ in the linear regime.
For stronger tilts, the downgoing Alfv\'en wave steepens to form an Alfv\'enic shock and the conversion efficiency saturates between 10\% and 20\%, however, the larger Poynting flux associated with stronger driving means the acoustic energy continues to increase. This strongly nonlinear regime produces the best candidates for observable sunquakes.

The second most important parameter is the magnetic field strength.
The triangles (green) in Fig.~\ref{fig:energies} show results for a {500~G} field and the squares (blue) show results for {1000~G}.  
For these cases $v_A=c_s$ at  $291\textrm{~km}$ and $117\textrm{~km}$ respectively.  For the parameters we have examined, the conversion efficiency is virtually independent of the field strength, however, larger Poynting flux in the Alfv\'enic front means that the energy in the acoustic wave increases significantly for stronger magnetic fields (scaling as $B_0^2$ in the ``weak tilt'' regime).

Changing the coronal field more rapidly increases the acoustic energy for weak angle changes but has a negligible effect in the strong regime that appears more relevant to observations. The dashed curve in Fig.~\ref{fig:energies} shows a 15~s driver for a 250~G field.  For the weak regime, $\mathbf{\delta J}$ in the Alfv\'enic front is inversely proportional to the vertical scale of the front, which is itself proportional to the duration of the driving, hence quicker drivers produce waves that are more nonlinear, increasing the coupling efficiency.  However, once the waves steepen to shocks, the impulse depends on the jump conditions across the magnetic shock and the acoustic energy becomes independent of the driver time scale.

Finally, we examine 1D simulations where the initial magnetic field strength varies with height, using the \citet{1983ZweibelHaber} scaling $B_0\propto P^\alpha$.  The diamonds (purple) in Fig.~\ref{fig:energies} show models with 250~G in the corona rising to 500~G at the photosphere, and the circles (black) have 500~G in the corona and 1000~G at the photosphere.  For weak tilts, models with varying $B_0(z)$ produce less acoustic energy than their uniform field counterparts with the same coronal field strength.  This is because the Alfv\'en speed ratio between the coupling height and corona is less extreme, giving less pile-up of magnetic flux in the magnetic front, smaller $\delta \mathbf{B}$ and $\delta \mathbf{J}$, smaller ponderomotive force and ultimately less efficient coupling. This effect is lost for large tilts.  In fact, in the strong regime, height variation of $B(z)$ assists generation of the acoustic wave because the Alfv\'enic shock dissipates less energy while propagating to the coupling height.

\section{Discussion}\label{sec:disc}
The MHD simulations in this paper show that changes of magnetic field direction are most acoustically active when the direction of a strong magnetic field changes by tens of degrees.
The 15 February 2011 X2.2 flare studied by \citet{2013Petrie} changed the photospheric magnetic angle by several tens of degrees, so while such changes are large, they have an observational basis for large flares.

The energy of the modeled acoustic wave becomes independent of the driving time scale when the magnetic change is large enough and fast enough that the Alfv\'enic front shocks before reaching the Alfv\'en-sound resonance, while for linear dynamics rapid magnetic changes are more effective at producing acoustic waves than gradual changes.  
\citet{2005SudolHarvey} fitted time series of GONG magnetogram data for 15 X-class flares and one third of the irreversible magnetic changes they identified had durations of less than 1 minute, which is the cadence of their observations.  We conclude that the driver durations used in our simulations (30 and 15~s) are plausible for the most dynamic locations. Considering longer time scales, half of the changes studied by \citet{2005SudolHarvey} had a duration of less than 1.5 minutes and three quarters had duration less than 10 minutes, although durations as long as several tens of minutes were also found therefore it may be relevant to some events that the acoustic flux does depend on driving time scale when sufficiently slow, even for large changes to the magnetic field.

The observation that would most naturally support our model, or rule it out for a given event by an unambiguous absence, is a suitable magnetic change near the acoustic source, consistent with the time the acoustic wave is launched. 
The idea of such a test is not new, however, our results justify some remarks.

Figure~\ref{fig:2d} indicates that the acoustic source should be horizontally aligned with the magnetic field change, consistent with the thesis of \citet{2009MartinezOliveros} and \citet{2012AlvaradoGomez} that these regions should be in close proximity. A small offset is expected when the magnetic field is inclined and the radiation providing the magnetic measurements comes from a different height to the acoustic source.  If chromospheric magnetic data are available, then a change there should preceed the photospheric change by the Alfv\'en travel time (as a guideline, several tens of seconds), which may help to distinguish magnetic changes due to coronal restructuring from other causes such as flux emergence.
We caution that magnetic changes may not show up in the line-of-sight magnetic field---the vertical magnetic field does not change at all in our simulations---or they may be missed in unresolved observations, e.g. if a localized change in the magnetic twist.  Thus, to be conclusive, such a test requires high-resolution vector measurements and careful treatment of the possibility of spatially unresolved magnetic changes.  

Continuing this theme, an interesting feature of our simulations is that the magnetic field undergoes a reversible change in addition to the irreversible change, producing a combined magnetic field evolution similar to the ones plotted for the seismic sources observed by \citet{2009MartinezOliveros} and \citet{2012AlvaradoGomez} (Fig. 3 in each paper). In our simulations, the reversible change is physical, the product of the flux of $B_z$ piling up in the Alfv\'en wave front when the Alfv\'en speed is significantly lower ahead of the wave front than behind it. There are good reasons to be cautious about short-lived features in magnetograms during flares in case they are spurious, which has led to sunquake generation by magnetic effects being assessed based on the irreversible component only \citep[e.g.][]{2012AlvaradoGomez}. However, our study indicates that the reversible part of magnetic signatures may actually correspond to a real reversible magnetic field perturbation. Consideration should therefore be given to the role of this reversible component in the energetics of sunquake excitation and whether it is possible that Lorentz drivers could be significant in events where they previously appeared to have been ruled out.

It should be emphasized as well that our simulations generate the acoustic wave by resonant coupling and the connection between magnetic field changes and acoustic waves is therefore more nuanced than a simple association between seismic sources and large changes of magnetic field or Lorentz force.  A given magnetic change may succeed or fail to produce a detectable sunquake depending on properties such as the degree of nonlinearity, the sound and Alfv\'en speed profiles, and the time scale.  These effects were not apparent from early theoretical works on this topic \citep{2008Hudson,2012Fisher} and it is our application of MHD theory that allows us to identify them now.  Future modeling may establish further properties that favor sunquake generation at particular locations, for example, the relative heights of Alfv\'en-sound resonance and the transition between optically thin and thick radiation may be significant when radiative damping is included.

The possibility of resonant coupling is also relevant to constraints that do not rely on observed field changes. A study of the main acoustic source in the 29 March 2014 X1 flare by \citet{2014Judge} reached an interesting set of conclusions, namely, that the sunquake power was at least two times greater than the downward enthalpy flux obtained using the Si~I line core, Poynting fluxes estimated using photospheric densities were only marginally sufficient for nonlinear perturbations, and several other forms of energy transport were ruled out.  How could the sunquake have been excited when it appears that no single transport mechanism operated through the full atmosphere?  The Si~I core typically forms between 200 and 500~km in 1D models \citep{2014Judge} and we would expect this to be above the $v_A=c_s$ surface for the reported field strength of 800~G.  Thus, if our mechanism were applied, the downward energy transport at the Si~I core heights would predominantly be as Poynting flux, which is very capable at these altitudes since the density is almost an order of magnitude lower than at the photosphere.  As energy approaches depths where Poynting flux cannot carry the required power, a portion is converted to an enthalpy flux to form the sunquake, but only below the altitudes sampled by the Si~I line.  Conversion of energy fluxes therefore resolves the apparent paradox. Combining MHD simulations and observations using techniques such as forward modeling should be a useful future partnership in this area.

It is pertinent at this point that our Fig.~\ref{fig:energies} (bottom) indicates a new and stronger constraint since at most 10\% to 20\% of the integrated Poynting flux into the acoustic kernel is converted to sunquake energy.  Interestingly, this percentage of the Poynting flux for a nonlinear magnetic front at the Si~I core height agrees well with the acoustic power per unit area for the 29 March 2014 sunquake, however, one should not read too much into this without supporting evidence of actual magnetic changes, which are currently unclear for this event (flux emergence at the time of the flare and near the acoustic source complicates interpretation of magnetic changes).  We also point out that since sunquake energy typically represents about 0.01--0.1\% of the total flare energy \citep{2011Donea}, one or more of the following must be true: the Poynting-flux energy that acoustically-active magnetic changes direct into the acoustic kernel is of order 0.1--1\% of the total flare energy (this does not include higher frequency waves of the sort considered by \citet{2016ReepRussell}), or the conversion efficiency actually attained is less than the maximum identified in this study, for example, due to dissipative processes that were not included in our simulations.

Future MHD modeling will add radiation, thermal conduction and resistivity, all of which may reduce the acoustic energy that ultimately enters the interior.  Radiation is particularly important because it modifies and damps acoustic waves in the lower atmosphere \citep{1985Fisher} and is therefore regarded as an obstacle to all sunquake mechanisms \citep{2012Fisher,2014Lindsey}. The Alfv\'en-sound resonance mechanism described in this paper appears  to have an advantage over particle beams in this regard because the Alfv\'en-sound resonance can potentially occur below the layers where acoustic waves are most strongly damped by radiation.  This applies primarily in regions of strong magnetic field strength such as sunspot penumbra, which is where sunquakes sources are almost exclusively observed. Future work should test this as a priority.
We also note that the relatively large perpendicular resistivity in the chromosphere \citep[][]{2013RussellFletcher,2014Leake} may dissipate some of the energy in the Alfv\'enic front before the acoustic wave is generated, thereby reducing the acoustic energy, and it remains to be shown whether or not these losses are significant.  

Another goal for future modeling is to address more general initial equilibria.  For example, sunquake sources are typically located in sunspot penumbra where the magnetic field is inclined from the outset, whereas we used an initially vertical magnetic field in this paper to simplify presentation of the theory and simulations. Similarly, the work presented did not include the geometrical effects of magnetic field convergence, which may alter $\delta B/B_0$ in the wave front and hence the efficiency of the resonant coupling.  Some magnetic field strengths and profiles may also move the $v_A=c_s$ resonance below the photosphere, where the gradient of $v_A/c_s$ at the resonance becomes gentler (Fig. \ref{fig:resonance}), which could in principle increase the efficiency of the resonant coupling by allowing it to occur over a larger interval.  Consideration of non-potential magnetic fields (for which $\mathbf{J}_0$ is non-zero) and non-static equilibria with background flows (such as the Evershed flow) would also be worthwhile.

Simulation studies based on localized regions are clearly valuable by themselves, however, future consideration should be given to the wider magnetic context of the flare as well.  One motivation for this is that the decrease in magnetic energy believed to power flares is a global phenomenon.  In the simulations presented here, localized shearing of the magnetic field by the applied driving increases the magnetic energy in the simulation domain by making the magnetic field more inclined.  This is known to occur in localized regions of the photosphere, with the local increases in magnetic energy more than balanced by decreases elsewhere in the active region \citep{2011Fletcher}.  A local reduction in the field inclination should also produce acoustic waves by the resonant coupling mechanism, as should changes to the magnetic azimuth, but some important physical differences mean that these scenarios should be explored in their own right.  Insights into the global pattern of magnetic changes in an active region during flares will be able to inform MHD studies of sunquake generation near particular features such as the polarity inversion line or flare ribbon hooks.  At the more ambitious end, since sunquake seismic sources have been observed coincident with the end points of an erupting flux rope \citep{2011Zharkov} it would be very interesting to explore this association by coupling a global MHD simulation of an erupting flux rope (similar, for instance, to \citet{2010Aulanier}) to a chromospheric model capable of capturing the resonant excitation.  Such an undertaking would be comparable in scale to present-day simulations of flux emergence and is therefore feasible with appropriate computing resources.

Alfv\'en wave fronts are only one method by which Lorentz forces may generate acoustic waves in the interior.
Another possibility is the transmission and mode conversion of magnetoacoustic waves originating in the corona, as considered analytically by \citet{2016Hansen}.  MHD simulations of the type presented here can be readily be adapted to the scenario they propose. Waveguided fast waves \citep{2013RussellStackhouse} and modes of the coronal structures (e.g. kink and sausage waves) would be interesting to consider in that context.

Finally, we point out two additional interesting features of our simulations.  The first is that the magnetic field change excites an oscillation that displaces the transition region with a period of 200~s, or approximately 3 minutes. This is evident in Fig.~\ref{fig:2d} and can be seen clearly in the online animation.  The simulated vertical velocity reaches up to a few tens of km/s for large magnetic changes; for comparison, IRIS \citep{2014IRIS} has a velocity resolution of 1~km/s. 
The other feature we highlight is the perpendicular current just below the photosphere at the end of our simulations (Fig.~\ref{fig:2d}(f)).  Coronal magnetic changes alter the coronal current system (our driver creates a pair of upward and downward field aligned currents) and since the current in the deep interior is unchanged, current continuity requires a new perpendicular current somewhere between these domains. The Alfv\'enic front naturally provides the required current closure \citep{1994WheatlandMelrose}, even when frozen by the vanishing Alfv\'en speed. The relatively large perpendicular resistivity at photospheric heights \citep{2013RussellFletcher,2014Leake} 
suggests the perpendicular current should heat the photosphere, potentially enhancing the optical emission where the tilt has changed -- a possibility that deserves investigation.

\section{Conclusions}

   \begin{enumerate}
      \item MHD simulations establish that changes to the coronal magnetic field excite acoustic waves in the solar interior with energy fluences matching or exceeding typical values for observed sunquakes.
      \item The acoustic wave is produced at the Alfv\'en-sound resonance in the lower atmosphere by the ponderomotive force in the Alfv\'enic front associated with the magnetic change.
      \item The most acoustically active changes to magnetic field direction are changes of tens of degrees on magnetic fields of hundreds or thousands of Gauss. 
   \end{enumerate}  
\

\begin{acknowledgements}
This work came out of the International Space Science Institute (Switzerland) Team 326 ``Magnetic Waves in Solar Flares'' led by AJBR and Lyndsay Fletcher (http://www.issibern.ch/teams/flarewaves). We thank ISSI for supporting the activity and the team members for useful discussions. We acknowledge grant ST/K000993/1 from STFC to the University of Dundee (AJBR) and summer student support from the School of Science \& Engineering (MKM). JEL was funded by NASA's Living With a Star Program and the Chief of Naval Research. The 2D simulation was run on the St Andrews MHD cluster (supported by STFC's Dirac programme) and analysed using VisIt (supported by the U.S. Department of Energy). We thank an anonymous referee for highly constructive comments.
\end{acknowledgements}

\end{document}